\documentclass[bibyear]{aa} 

\usepackage{txfonts}
\usepackage{hyperref}

\begin{document}

   \title{Eigenmode analysis of perturbations in the primordial medium at and before recombination}

\titlerunning {Eigenmode analysis of Perturbations in the Primordial Medium}
  
   \author{A. H. Nelson }

   \institute{School of Physics and Astronomy, Cardiff University,
             Cardiff CF24 3AA, United Kingdom\\
              \email{nelsona@cf.ac.uk}
             }

   \date{Received December 1, 2020}

  \abstract
    {Anisotropies of the cosmic microwave background are thought to be due to perturbations of the primordial medium, which, post recombination, lead to the formation of galaxy clusters and galaxies}
   {The perturbation wave modes of the primordial medium at and before recombination, consisting of a fully\ ionised\ baryonic plasma, a strong black body radiation field, and cold dark matter, are analysed.}
{We use the linear perturbation theory of the relativistic equations of motion, utilising a strict thermodynamic equilibrium model that relates the radiation energy density to the plasma temperature. }
{   It is shown that a wave mode corresponding to the postulated baryon\ acoustic waves exists with a phase velocity close to the speed of light, but the participation of the dark matter in this mode is very small.  Instead, the dark matter has its own dominant mode in the form of\ gravitational\ collapse,\ with very little participation by the baryonic plasma. }
{  \ In view of this very weak coupling between baryons and dark matter, the initial conditions postulated for computer simulations of large-scale structure and galaxy formation -- which assume that after recombination, when galaxy formation is getting underway, baryon and dark matter density perturbations are spatially coincident and are equal in terms of fractional amplitude -- may be unjustified.\  Additionally, the possible non-coincidence of baryon and dark matter perturbations at the last scattering surface has implications for the analysis of cosmic microwave background anisotropies.   }


\maketitle
%

\section{Introduction}

\vspace{\baselineskip}
The accepted orthodoxy in cosmology is that the Universe prior to a redshift of 1500\  consisted partly of fully ionised baryonic plasma at a temperature above 4000\textsuperscript{o}K\ in\ thermal\ equilibrium,\ a consequent black body radiation field with a radiation pressure far exceeding the thermal pressure of the plasma, and a dark matter component with a mass an order of magnitude greater than that of the baryons.    Due to the radiation pressure acting on the baryons,  the fluctuations in baryon density, inferred from the anisotropies in the cosmic microwave background (CMB) radiation, would be unable to collapse to form galaxies and galaxy clusters, since the pressure was such that the Jeans length (i.e. the minimum scale for a density enhancement to collapse under gravity) would be greater than the scale of the Universe.

Subsequent to redshift 1500, the temperature of the  expanding plasma would fall to below the temperature needed for hydrogen recombination to occur in thermal equilibrium, and baryonic matter would then decouple from the radiation field after this era of recombination.   Consequently, radiation pressure would no longer prevent the baryon density perturbations from collapsing under gravity, and the era of galaxy and cluster formation would begin.

Large-scale cosmological computer simulations start their initial conditions after the era of  full recombination at redshift $ \sim $ 100 and assume that there is coincidence between dark matter and the baryonic perturbations that are revealed by the CMB anisotropies; in other words, they assume that the positions and amplitudes of peaks and troughs of density perturbations in terms of a percentage of the ambient density are identical for baryonic and dark matter (Vogelsberger\ et al  2014; Schaye et al 2015; Kaviraj et al 2017 ). However, there is a serious issue with this assumption: though we do not know the specific nature of dark matter, it seems certain that it does not absorb or emit electromagnetic radiation and hence cannot be affected at all by the radiation pressure.  This means that prior to the era of recombination, perturbations of dark matter cannot be supported by radiation pressure, while baryonic perturbations will be supported. Furthermore, standard cosmology postulates cold dark matter (\( \Lambda \)CDM), so there is also no internal kinetic energy that would be able to prevent the collapse.  Therefore, dark matter should be subject to gravitational collapse on all scales.\  This would leave any assumption of coincidence at recombination between baryonic matter perturbations, as evidenced by the electron density perturbations deduced from the CMB, and dark matter perturbations open to question.\\

\vspace{\baselineskip}
To\ quantify\ this idea, it is useful  to carry out a simple plane wave analysis of the perturbation of a system consisting of baryonic plasma, black body radiation, and dark matter.  This analysis   considers the period between horizon crossing of the perturbation wavelength and recombination.   A similar analysis has been the basis of the baryon acoustic model for the interpretation of the CMB, as exemplified by Dodelson (2003). In this paper we follow a similar physical model for the baryons and the dark matter but use a different physical paradigm for the radiation.  Dodelson uses a kinetic theory model for the photons of the radiation field, in the form of the Einstein Boltzmann equations, leading to a hierarchy of moment equations for the radiation temperature perturbation.\  We assume strict thermodynamic equilibrium, with the radiation having a black body spectrum and the energy density given by the Stefan-Boltzmann law based on the baryon temperature.

It should be noted  that  the fully ionised baryon plasma does not support simple acoustic waves, but, in common with contemporary plasmas, it supports a variety of waves, such as Langmuir waves (also known as plasma waves, where the electrons oscillate rapidly, independent of the ions), ion-acoustic waves (Boyd and Sanderson 1969), and, if a magnetic field is present in the primordial plasma (Ruis-Granados, Battaner $\&$  Florido 2016),\ a  variety of magnetohydrodynamic waves.  However, ion acoustic waves are the nearest mode to a simple acoustic wave, and the plasma frequency at recombination is many magnitudes higher than the frequency of ion acoustic waves at the relevant wavelengths, so the electrostatic binding of electrons and ions is very strong. Therefore,  in common with Dodelson’s analysis, we assume that a simple acoustic-like wave is the origin of the CMB anisotropies and follow a similar set of equations for baryons and dark matter as that on page 111 of Dodelson (2003).

In this paper we perform an eigenmode analysis of the plane waves supported by this model.\ This analysis highlights the very weak coupling between the baryon and dark matter perturbations and calls into question the assumptions behind the initial conditions for simulations starting at redshift$ \sim $ 100, and indeed the analysis of the CMB anisotropies in terms of coincident baryon and dark matter perturbations.

\section{Basic equations with black body radiation pressure}

Our assumption that the radiation is in strict thermodynamic equilibrium with the baryonic plasma is based on the fact that we considered only scales much greater than the mean free path of a photon, meaning that the radiation\ and\ plasma move together; in other words, we used the tight coupling approximation. This neglects the effect of Silk damping at short wavelengths  (Silk 1967) but yields a simple, but plausible, physical model that identifies the different baryon and dark matter wave modes. The energy density of the radiation is then given by the Stefan-Boltzmann law, leading to the associated isotropic, but spatially varying, radiation\ pressure.

We note that under the  tight coupling approximation, discussed by Dodelson (2003) on page 226, the radiation dipole term in the equation of motion of the baryons in his formalism, proportional to             \(\Theta _1\), can be shown, using the equation of motion for \(\Theta _1\), to reduce to a gradient of the monopole term (i.e. \(i k \)\(\Theta _0\)) when the frequency of a wave is much lower than the photon collision frequency.  Consequently, under these assumptions, the two radiation formalisms are equivalent. 

The total thermal energy density is then

 \( U= U_{R~}+U_{B}\),

\noindent where

 \( U_{B}= \frac{3}{2}n_{B~}KT = 3 \frac{ \rho _{B}}{m_{p}}KT. \)  

\noindent Here,  \(  \rho _{B} \)  is the density of baryons,  \( m_{p} \)  is the proton mass, T is the baryonic temperature, and K is Boltzmann’s constant, while 

 \( U_{R}=~ 4 \frac{ \sigma T^{4}}{c}\),

\noindent where $ \sigma $  is Stefan’s constant, and c is the speed of light.  

 With T = 4000\textsuperscript{o}K,\  and   \(  \rho _{B} \sim 10^{-17} \) kg/m\textsuperscript{3\ }  at z = 1500

\noindent we have  \( U_{R~} \gg U_{B}.\)

Hence, we deemed the work done by radiation pressure on baryonic compressions to go wholly into  \( U_{R} \)  since when T increases radiation will be emitted, increasing   \( U_{R} \). We therefore used a conservation equation for the  \( U_{R} \)  effectively carried by the baryons, with the addition of a source term that represents adiabatic compressional heating together with a relativistic gravitational correction term.

The perturbed, expanding flat Universe metric in the Newtonian conformal gauge is given by
 \[ \text{~~~~~~~~~~~~~~~~~~~~~~ g}_{00} \left(\textbf{ x},t \right) =-1-2 \psi  \left( \textbf{x},t \right)  \]

 \[ ~~~~~~~~~~~~~~~~~~~~~~~~g_{0i} \left(\textbf{ x},t \right) =0 \] 

 \[ \text{~~~~~~~~~~~~~~~~~~~~~~~~ g}_{ij} \left(\textbf{ x},t \right) =a \left( t \right) ^{2} \delta _{ij} \left( 1+2 \phi  \left( \textbf{x},t \right) \right), \] 

\noindent where  \( a \left( t \right)  \)  is the cosmological scale factor, \({ \psi} \)  is the Newtonian potential, and  \({ \phi} \)  is the perturbation of the spatial curvature.   We ignored any neutrino contribution since such a weakly interacting component will have little effect on the relationship between baryons and dark matter, and, since the radiation field is isotropic, we have  (see Eq. (5.33) of Dodelson 2003)\[  ~~~~~ ~~ \phi =- \psi.  \]

\noindent Hence, the set of equations governing a combined baryonic and dark matter fluid accompanied by a black body radiation field with dominant radiation pressure, replacing Eqs. (4.100) to (4.106) of Dodelson, is

\vspace{\baselineskip}
 \( ~~~~~~~~~~~~~~~~~~~~~~~~~~~~~~\frac{ \partial \rho _{B}}{ \partial t}+ \triangledown . \left(  \rho _{B}\textbf{v}_{B} \right) ~ = -\frac{3}{c^{2}}~ \rho _{B}\frac{ \partial  \phi }{ \partial t}~~~~~~~~~~~~~~~~\left( 1 \right)  \)  \\

 \(   \rho _{B} \left( \frac{ \partial }{ \partial t}+ \textbf{v}_{B}.\triangledown  \right) \textbf{v}_{B}+ \triangledown p + \frac{da/dt}{a} \rho _{B}\textbf{v}_{B}= - \)   \(  \rho _{B}\triangledown  \phi  \) \ \ \ \ \ \ \ \ \ \ \ \ \ \ \ \ \ \ (2)\\

 \(~~~~~~~~~~~~~~~~~~~~~~~~~~~~ \frac{ \partial  \rho _{D}}{ \partial t}+ \triangledown . \left(  \rho _{D}\textbf{v}_{D} \right) ~ = -\frac{3}{c^{2}}~ \rho _{D}\frac{ \partial  \phi }{ \partial t}~~~~~~~~~~~~~~~~  \left( 3 \right)  \)  \\

 \(  ~~~~~~ \rho _{D} \left( \frac{ \partial }{ \partial t}+ \textbf{v}_{D}.\triangledown  \right) \textbf{v}_{D} + \frac{da/dt}{a} \rho _{D}\textbf{v}_{D}= - \)   \(  \rho _{D}\triangledown  \phi  \) \ \ \ \ \ \ \ \ \ \ \ \ \ \ \ \ \  \ \ \  (4)\\

 \(~~~~~~~~~~~~~~~ \frac{ \partial U_{R}}{ \partial t}+ \triangledown . \left( U_{R}\textbf{v}_{B} \right) +p\triangledown .\textbf{v}_{B}= -\frac{3}{c^{2}}~U_{R}\frac{ \partial  \phi }{ \partial t}~~~~~~~~~~~~~~~~ \left( 5 \right)  \)  \\

\noindent and finally\\

 \(~~~~~~~~~~~~~~~~~~~~  \triangledown ^{2} \phi - \frac{3}{c^{2}}\frac{\frac{da}{dt}}{a} \left( \frac{ \partial  \phi }{ \partial t}+\frac{\frac{da}{dt}}{a} \phi  \right)  =4 \pi G \left(    \rho _{D}+  \rho _{B}+ \frac{U_{R}}{c^{2}} \right).  \)    (6)\\

\noindent Here,  \( ~\textbf{v} \) \textbf{ }is\ velocity  and\\

 \( p= \frac{4}{3}~\frac{ \sigma T^{4}}{c}~  \gg  n_{B~}KT. \)

\section{Plane wave perturbation solutions with frequency >>  \( \frac{da/dt}{a}~~ \)}

We firstly ignore the rate of change of the cosmological scale factor terms, under the assumption that we are considering only perturbations with timescales much smaller than the expansion time, so the expansion rate is negligible compared to the rate of change of the perturbations. The terms proportional to  \( \frac{ \partial  \phi }{ \partial t} \)  on the right-hand side of the conservation equations, Eqs. (1), (3), and (5), are the only relativistic correction terms required under this assumption and will only be non-negligible in the case of a wave mode where the phase velocity is comparable to c.

Linearising with respect to the small perturbations of the principal physical quantities (i.e.  \( \overline{ \rho }_{B} \), \( ~\overline{\textbf{v}}_{B} \), \( ~\overline{ \rho }_{D} \),   \( ~\overline{\textbf{v}}_{D} \), \( ~\overline{ \phi } \), and \(\overline{T} \)) and then assuming longitudinal plane waves propagating in the  \( x \) \  direction (i.e. all perturbations \( \propto exp\left(i \omega t -ikx \right)  \)  with  \( \overline{\textbf{v}}_{B,D}= \left( \overline{v}_{B,D},o,o \right)  \)), we have from Eqs. (1) to (6)\\

 \(~~~~~~~~~~~~  i \omega ~\overline{ \rho }_{B}- ik \)   \(  \rho _{B}\overline{v}_{B}= -\frac{3}{c^{2}} \rho _{B}i \omega \overline{ \phi }~~~~~~~~~~~~~~~~~~~~~~~~~~~~~~~~  \left( 7 \right)  \) \\

 \(~~~~  i \omega  \rho _{B}\overline{v}_{B}~- ik \)   \( \frac{16 \sigma T^{3}}{3c}\overline{T}=  \rho _{B}ik\overline{ \phi }~~~~~~~~~~~~~~~~~~~~~~~~~~~~~~~~~~~~~~~~  \left( 8 \right)  \) \\

 \(~~~~~~~~~~~  i \omega ~\overline{ \rho }_{D}- ik \)   \(  \rho _{D}\overline{v}_{D}= -\frac{3}{c^{2}} \rho _{D}i \omega \overline{ \phi }~~~~~~~~~~~~~~~~~~~~~~~~~~~~~~~~~  \left( 9 \right)  \) \\

   \(~~~~~~~~~~~~~~~~~~~~~~~ i \omega  \rho _{D}\overline{v}_{D}~=  \rho _{D}ik\overline{ \phi }~~~~~~~~~~~~~~~~~~~~~~~~~~~~~~~~~~~~~~  \left( 10 \right)  \)  \\

 \(  i \omega ~\frac{16 \sigma T^{3}}{c}\overline{T}- ik\frac{16 \sigma T^{4}}{3c} \)   \( \overline{v}_{B}= -\frac{12 \sigma T^{4}}{c^{3}}i \omega \overline{ \phi }~~~~~~~~~~~~~~~~~~~~~~~~~~~~~~~~  \left( 11 \right)  \) \\

 \(~~~~~~~~~~~~~~~~~~~~~~~~~~~~~~~~~~ \text{ k}^{2}\overline{ \phi } =-4 \pi G \left(  \overline{ \rho }_{D}+ \overline{ \rho }_{B}+ \frac{16 \sigma T^{3}}{c^{3}}\overline{T} \right).   \ \ \ \ \ \ \   (12)  \) \\

\vspace{\baselineskip}
We need to now reduce these equations to a single equation for one of the six physical quantities in which the multiplying expression will yield a dispersion relation for frequency,  \(  \omega  \),  as a function of wavenumber,  \( k \) .\\

From Eq. (10),\ \  \ \ \ \ \ \ \ \ \ \ \ \ \ \ \ \ \ \ \ \  \ \ \ \ \ \ \ \ \( \overline{ \phi }= \frac{ \omega }{k}~\overline{v}_{D~}, \)  \ \ \ \ \ \ \ \ \ \ \ \ \  \ \ \ \ \ \ \ \ \ \  \ \ \ \ (13)\\

\noindent and\ from Eq.\ (7),\ \ \  \ \ \ \ \ \ \ \ \ \ \ \ \ \ \ \ \ \ \ \ \ \ \    \( \overline{v}_{B}= \frac{ \omega }{k}~\frac{\overline{ \rho }_{B}}{ \rho _{B}}+ \frac{3}{c^{2}}~\frac{ \omega }{k} \) \   \( \overline{ \phi, } \) \  \ \ \ \ \ \ \ \ \ \ \ \  (14)\\

\noindent while\ from\ Eq. (11)\  \ \ \ \ \ \ \ \ \ \ \ \  \ \ \ \ \ \ \ \    \( \overline{T}= \frac{k}{3 \omega }T\overline{v}_{B}- \frac{3}{4c^{2}}T\overline{ \phi. } \) \\

 \noindent Using Eq. (14), this becomes\ \   \( \overline{T} \)   \( = \frac{T}{3}\frac{\overline{ \rho }_{B}}{ \rho _{B}}+ \frac{1}{4c^{2}}T\overline{ \phi. } \)  \   \ \ \ \ \ \ \ \ \ \ \ \ \ \   (15)\\

\noindent Therefore, substituting for  \( \overline{T} \) \ in Eq. (8)  we have\\

  \(  \omega  \rho _{B}\overline{v}_{B}- k\frac{16 \sigma T^{4}}{9c}\frac{\overline{ \rho }_{B}}{ \rho _{B}}- k\frac{4 \sigma T^{4}}{3c^{3}}\overline{ \phi }=  \)   \(  \rho _{B}k \)   \( \overline{ \phi, } \) \ \ \ \ \ \ \  \\

\noindent and using Eq. (14) we have\\

 \(  \ \ \ \ \ \ \ \ \ \  \left( \frac{ \omega ^{2}}{k^{2}}-  \alpha c^{2} \right) \overline{ \rho }_{B}=- \left(  \frac{3}{c^{2}}~\frac{ \omega ^{2}}{k^{2}}-1- \frac{3}{4} \alpha  \right)   \rho _{B}\overline{  \phi, } \) \ \  \ \ \ \ \ \ \ \  \ \ \ \ \ \   (16)\\

\noindent where\ \   \(  \alpha = \frac{16 \sigma T^{4}}{9 \rho _{B}c^{3}}. \) \ \  \\

\noindent From\ Eq. (9),\   \ \ \ \ \ \ \ \  \ \ \ \ \  \  \( \overline{ \rho }_{D}= \frac{k}{ \omega } \rho _{D}\overline{v}_{D}- \frac{3}{c^{2}}~ \rho _{D}\overline{ \phi } \) \ \ \ \ \  \ \ \ \ \ \ \ \ \ \ \ \ \ \ \ \ \ \ \ \ \ \ \ \ \ \ \   \    (17)\\

\noindent and,\ using\ Eq. (13),\  \ \    \( \overline{ \rho }_{D}= \left( \frac{k^{2}}{ \omega ^{2}}- \frac{3}{c^{2}}  \right)  \rho _{D}\overline{ \phi. } \) \ \ \ \   \ \ \ \ \ \ \  \ \ \ \ \ \ \   \ \ \ \ \ \ \   \ \ \ \ \ \  (18)\\

\noindent From Eq. (15), Eq. (12) becomes\\

 \(  k^{2}\overline{ \phi } =-4 \pi G \left( \overline{ \rho }_{D}+ \overline{ \rho }_{B} \left( 1+3 \alpha  \right) + \frac{4 \sigma T^{4}}{c^{5}}\overline{ \phi } \right).  \) \ \ \ \ \ \ \ \ \ \ \ \  \\

\noindent Therefore, substituting for  \( \overline{ \rho }_{B} \)  and  \( \overline{ \rho }_{D} \)  from Eqs. (16) and (18), respectively, we obtain\\

 \(  \{ k^{2}+4 \pi G \left( \frac{k^{2}}{ \omega ^{2}}- \frac{3}{c^{2}}~ \right)  \rho _{D}-4 \pi G \left( 1+3 \alpha  \right) \frac{ \left( \frac{3}{c^{2}}~\frac{ \omega ^{2}}{k^{2}}- 1_{~}-~ \frac{3 \alpha }{4} \right) }{ \left( \frac{ \omega ^{2}}{k^{2}}-  \alpha c^{2} \right) } \rho _{B}\)
\\\\\\
\vspace{\baselineskip}
\(~~~~~~~~~~~~~~~~~~~~~~~~~~~~~~~~~~~~~~~~~~~~~~~~~~~~~~~~~~~~~~~~~~~~~~~+\frac{9 \pi G \alpha  \rho _{B}}{c^{2}} \}  \)   \( \overline{ \phi } \)  = 0.\\

The\ expression\ inside the curly brackets must therefore be zero and represents the dispersion relation for the perturbations.   Multiplying it by  \(  \omega ^{2} \left( \frac{ \omega ^{2}}{k^{2}}-  \alpha c^{2} \right) ~ \) and defining the characteristic frequencies (which are actually inverse timescales),\\

 \(  \omega _{D}^{2}=  \)   \( 4 \pi G \rho _{D} \) \ \ and\    \(  \omega _{B}^{2}=  \)   \( 4 \pi G \rho _{B} \). \\

\noindent We then obtain the following quadratic equation for  \(  \omega ^{2} \): \\

 \(  \left[ 1-\frac{3 \omega _{D}^{2}}{k^{2}c^{2}}-3 \left( 1+  \frac{9 \alpha }{4} \right) \frac{ \omega _{B}^{2}}{k^{2}c^{2}} \right]   \omega ^{4}\) \( - \left[   \alpha k^{2}c^{2}- \left( 1+3 \alpha  \right)  \omega _{D}^{2}- \left( 1+\frac{15 \alpha }{4} \right)  \omega _{B}^{2} \right]  \omega ^{2}- \alpha k^{2}c^{2} \omega _{D}^{2}=0. \) \\

\noindent This has the solutions\\

 \(  \omega ^{2}= \{ \frac{ \left[  1- \left( 1+3 \alpha  \right) \frac{ \omega _{D}^{2}}{ \alpha k^{2}c^{2}}- \left( 1+\frac{15 \alpha }{4} \right) \frac{ \omega _{B}^{2}}{ \alpha k^{2}c^{2}} \right] 
 \pm \sqrt[]{A + B}}{2 \left[ 1-\frac{3 \omega _{D}^{2}}{k^{2}c^{2}}-3 \left( 1+~\frac{9 \alpha }{4} \right) \frac{ \omega _{B}^{2}}{k^{2}c^{2}} \right] } \}  \alpha k^{2}c^{2}, \)  \  \   (19)\\\\

\vspace{\baselineskip}
\noindent where\\

 A=\(\left[  1- \left( 1+3 \alpha  \right) \frac{ \omega _{D}^{2}}{ \alpha k^{2}c^{2}}- \left( 1+\frac{15 \alpha }{4} \right) \frac{ \omega _{B}^{2}}{ \alpha k^{2}c^{2}} \right] ^{2}\)

\vspace{\baselineskip}
\noindent and\\

B=\(\frac{4 \omega _{D}^{2}}{ \alpha k^{2}c^{2}} \left[ 1-\frac{3 \omega _{D}^{2}}{k^{2}c^{2}}-3 \left( 1+~\frac{9 \alpha }{4} \right) \frac{ \omega _{B}^{2}}{k^{2}c^{2}} \right]. \)

\vspace{\baselineskip}
Now at z = 1500,\   \(  \rho _{B} \sim 10^{-17} \) kg/m\textsuperscript{3}\ \ and\    \(  \rho _{D} \sim 10^{-16} \) kg/m\textsuperscript{3}; therefore,  \(  \omega _{B}^{2} \sim 10^{-26} \)  and  \(  \omega _{D}^{2} \sim 10^{-24} \). For wavelengths $ \sim $  10 Kpc at recombination (and therefore within the horizon),  \( k^{2}c^{2} \sim 10^{-22} \).\\

\noindent Therefore, both  \( \frac{ \omega _{D}^{2}}{k^{2}c^{2}} \)  and  \( \frac{ \omega _{B}^{2}}{k^{2}c^{2}}  \ll 1 \),  \\

\noindent and, to first order in these small quantities, the argument of the square root here becomes\\

A+B= \( 1+2 \left( 1-3 \alpha  \right)  \)   \( \frac{ \omega _{D}^{2}}{ \alpha k^{2}c^{2}}-  \) 2 \(  \left( 1+\frac{15 \alpha }{4} \right) \frac{ \omega _{B}^{2}}{ \alpha k^{2}c^{2}} \), \\

\noindent and the top line of the fraction becomes\\

 \( 1- \left( 1+3 \alpha  \right)  \)   \( \frac{ \omega _{D}^{2}}{ \alpha k^{2}c^{2}}-  \left( 1+\frac{15 \alpha }{4} \right) \frac{ \omega _{B}^{2}}{ \alpha k^{2}c^{2}} \)\\
\vspace{\baselineskip}
\ \ \ \ \ \ \ \ \ \ \ \ \ \ \ \ \ \ \  \ \ \ \ \ \ \ \ \ \( \pm  \left[ 1+ \left( 1-3 \alpha  \right)   \frac{ \omega _{D}^{2}}{ \alpha k^{2}c^{2}}-  \left( 1+\frac{15 \alpha }{4} \right) \frac{ \omega _{B}^{2}}{ \alpha k^{2}c^{2}} \right].  \) 

\noindent For the minus sign, this reduces to  \( -2\frac{ \omega _{D}^{2}}{ \alpha k^{2}c^{2}} \), meaning the solution for  \(  \omega ^{2} \) \  is

 \[  ~~~~~ ~~ \omega ^{2}= - \omega _{D}^{2}. \]

\noindent Therefore, the time dependence of all perturbations is  \( \exp   \left(  \pm  \omega _{D}t \right)  \) \ (i.e. there is one decaying mode and one unstable mode).   For both these modes, Eqs. (16) and (18) imply \\

 \(  \left( \frac{ \omega _{D}^{2}}{k^{2}}+  \alpha c^{2} \right) \overline{ \rho }_{B}=- \left( ~\frac{3 \omega _{D}^{2}}{c^{2}k^{2}}+1+~ \frac{3}{4} \alpha  \right)  \rho _{B~}\overline{  \phi } \) \ \ \  \\

\noindent and\\

 \( \overline{ \rho }_{D}=- \left( \frac{k^{2}}{ \omega _{D}^{2}}+ \frac{3}{c^{2}}  \right)  \rho _{D}\overline{ \phi. } \) \ \ \  \\

\noindent Therefore, \\

 \( \frac{\overline{ \rho }_{D}}{\overline{ \rho }_{B}} = \frac{1+\frac{ \alpha k^{2}c^{2}}{ \omega _{D}^{2}}+\frac{3 \omega _{D}^{2}}{c^{2}k^{2}}+3 \alpha }{1+\frac{3 \omega _{D}^{2}}{c^{2}k^{2}}+\frac{3 \alpha }{4}}\frac{ \rho _{D}}{ \rho _{B~}}. \)  \\

\noindent The dominant term here is the second one on the top line, and hence\\

 \( \frac{\overline{ \rho }_{D}}{\overline{ \rho }_{B}}  \approx  \frac{ \alpha k^{2}c^{2}}{ \omega _{D}^{2} \left( 1+\frac{3 \alpha }{4} \right) }~\frac{ \rho _{D}}{ \rho _{B~}} \gg 1 \),  \\

\noindent that is, the dark matter and baryonic matter density perturbations are in phase but with the dark matter perturbation very much greater than the\ baryonic\ perturbation.   Although the baryonic matter reacts to the potential well created by the dark matter, radiation pressure prevents the baryonic matter from falling significantly into that well.  Now from Eq. (17), and substituting for  \( \overline{ \phi } \)  from Eq. (18), we obtain\\

 \( \overline{v}_{D}=\frac{k}{ \omega  \left( \frac{k^{2}}{ \omega ^{2}}- \frac{3}{c^{2}}  \right) }\frac{\overline{ \rho }_{D}}{ \rho _{D}} \),  \ \ \ \  \\

\noindent which for these two modes yields\\

 \( \overline{v}_{D}= \pm \frac{ik}{ \omega _{D} \left( \frac{k^{2}}{ \omega _{D}^{2}}+ \frac{3}{c^{2}}  \right) }\frac{\overline{ \rho }_{D}}{ \rho _{D}}, \) \ \ \ \ \  \\

\noindent meaning the velocity perturbation is 90\textsuperscript{o} out of phase with the density perturbation and, for the decaying mode (shown with the plus sign), involves expansion of the dark matter away from the density peak. On the other hand, the unstable mode (minus\ sign)\ involves\ gravitational collapse into the density peak. Any initial dark matter perturbation will potentially involve both of these modes; for instance, an initial density perturbation at rest involves equal amplitude contributions for both modes – one mode decays away, while the other mode grows exponentially in amplitude.   The cold dark matter is therefore unconditionally unstable to gravitational collapse on all scales.\\

For the plus sign the solution (19) for \(  \omega ^{2} \)  to first order in  \( \frac{ \omega _{D}^{2}}{k^{2}c^{2}} \) \  and  \( \frac{ \omega _{B}^{2}}{k^{2}c^{2}} \)  is\\

 \(  \omega ^{2} \) \ \ =   \(  \alpha k^{2}c^{2}+  \left( \frac{27}{4} \alpha ^{2}-\frac{3}{4} \alpha -1 \right)  \omega _{B}^{2}  \approx   \alpha k^{2}c^{2}. \) \\

\noindent Again, for these modes, Eqs. (16) and (18) imply\\

 \(  \left( \frac{27}{4} \alpha ^{2}-\frac{3}{4} \alpha -1 \right) \frac{ \omega _{B}^{2}}{k^{2}}\overline{ \rho }_{B}=    \left( 1-\frac{9}{4} \alpha - 3 \left( \frac{27}{4} \alpha ^{2}-\frac{3}{4} \alpha -1 \right) \frac{ \omega _{B}^{2}}{k^{2}c^{2}} \right)   \rho _{B}\overline{  \phi } \) \  \\

\noindent and\\

\ \   \( \overline{ \rho }_{D}= \left( \frac{k^{2}}{ \alpha k^{2}c^{2}+  \left( \frac{27}{4} \alpha ^{2}-\frac{3}{4} \alpha -1 \right)  \omega _{B}^{2}}-\frac{3}{c^{2}}  \right)  \rho _{D}\overline{ \phi } \).\\

\noindent That is,\  to first order,\\

 \( \frac{\overline{ \rho }_{D}}{\overline{ \rho }_{B}} \)   =  \( \frac{ \left( 1-3 \alpha  \right)  \left( \frac{27}{4} \alpha ^{2}-\frac{3}{4} \alpha -1 \right) }{ \alpha  \left( 1- \frac{9}{4} \alpha  \right) }~~\frac{ \omega _{B}^{2}}{k^{2}c^{2}}\frac{ \rho _{D}}{ \rho _{B}} \) \  << 1\\

\noindent at the era of recombination  \(  \alpha   \approx 0.098 \),  and hence\\

 \[ ~~~~~ \frac{\overline{ \rho }_{D}}{\overline{ \rho }_{B}}  \approx 9.29\frac{ \omega _{B}^{2}}{k^{2}c^{2}}\frac{ \rho _{D}}{ \rho _{B}}, \] \\

\noindent meaning  \( \overline{ \rho }_{D} \)  is in phase with  \( \overline{ \rho }_{B} \)   but is very small in comparison. This mode is the postulated baryon acoustic oscillations, with a phase velocity close to  \( 0.3c \)  due to being driven by radiation pressure rather than baryonic thermal pressure.

For the dark matter gravitational collapse mode, the assumption that  \(  \omega  \)  >>   \( \frac{da/dt}{a} \) \  turns out to be untenable, and hence neglecting the scale factor terms is not justifiable -- prior to recombination there would only be a few\ e-folding times for the gravitational collapse to proceed.  However, for the baryon acoustic oscillations, for wavelengths $ \sim $  10 Kpc at recombination (and therefore within the horizon), \\

 \(  \omega  \sim kc  \)  >>   \( \frac{da/dt}{a} \) , \\ so the analysis holds\textbf{ }for baryon acoustic\ oscillations.

\section{Analysis including the scale factor terms for the dark matter collapse perturbations}

We sought a solution of Eqs. (1)-(6) where the perturbations of\( ~  \rho _{B},~v_{B} \), and  \( U_{R} \) \ are\ negligible, but we retained the scale factor terms in Eq. (4).   We can still ignore the scale factor terms in Eq. (6) since  \( k^{2}c^{2} \gg \frac{da/dt}{a} \omega _{D} \)  and  \(  \left( \frac{\frac{da}{dt}}{a} \right) ^{2} \). We can also drop the  \( \frac{ \partial  \phi }{ \partial t} \)  term on the right-hand side of Eq. (3) since we expect the characteristic timescale to be >>  \( 1/kc \) .

Equations (3), (4), and (6) for the Fourier component \( exp \left(-ikx \right) \) then become\\

 \( ~~~~~~~~~~~~~~~~~\frac{d\overline{ \rho }_{D}}{dt} = ik \)   \(  \rho _{D}\overline{v}_{D}, \)  ~~~~~ ~~~~~ ~~~~~~~~~~~~~~~~~~~~~~~~~~~ (20)\\

 \(   \frac{~d\overline{v}_{D}}{dt} + \frac{da/dt}{a}\overline{v}_{D}= ik \)   \( \overline{  \phi, } \) ~~~~~~ ~~~~~ ~~~~~~~~~~~~~~~~~~~~~~~~~~~~~~~~~ (21)\\

\noindent and  \( ~~~~~~~~~~~~~~~~ k^{2}\overline{ \phi } =-4 \pi G \)   \( \overline{ \rho }_{D}. \) ~~~~~~ ~~~~~ ~~~~~~~~~~~~~~~~~~~~~~~~~(22)\\

\noindent Combining Eqs. (21) and (22) and multiplying both sides by   \( a \),  we have\\

    \ \ \ \ \ \ \   \( \frac{ d \left( a\overline{v}_{D} \right) }{dt} = -i \)   \( \frac{4 \pi Ga}{k}\overline{ \rho }_{D}. \) \\

\noindent This yields the set of equations\\

 \ \ \ \ \ \ \ \(\frac{d\overline{ \rho }_{D}}{dt}~~=  \frac{k \rho _{D}}{a}\overline{u}_{D}~~~~~~~~~~~~~~~~~~~~~~~~~~~~~~~~~~~~~~~~~~~ \) \\

 \noindent and\ \ \ \ \ \ \ \  \( ~\frac{d\overline{u}_{D}}{dt}=  \)   \( \frac{4 \pi Ga}{k}\overline{ \rho }_{D}, \) \ \ \ \ \ \ \ \ \ \ \ \ \ \ \ \ \ \ \ \ \ \ \ \ \ \ \ \ \ \ \ \ \ \ \ \ \ \ \ \ \ \ \ \ \ \ \ \ \  \\

\noindent where  \( ~ia\overline{v}_{D}~~=  \overline{u}_{D} \) .\\

Solving these numerically, starting with  \( \frac{\overline{ \rho }_{D}}{ \rho _{D}}= 10^{-6} \)  and  \( \overline{u}_{D}=0  \) at \( a~ = 10^{-5}\), \ \  \( \overline{ \rho }_{D} \)  increases almost linearly by a factor of just under 3 at \( a~ = 10^{-3}\), corresponding to recombination. This solution has a potential field given by

 \[  ~~~~~ ~~~\overline{ \phi }~ = -\frac{4 \pi G}{k^{2}}\overline{ \rho }_{D}. \]

\noindent This can be used to estimate the response of the baryons driven by this dark matter potential by equating the radiation pressure, which will oppose the collapse of the baryons to the gravitational pull of the  potential well, yielding

 \[ ~~~~~~\frac{16 \sigma T^{3}}{3c}\overline{T} = - \rho _{B}\overline{ \phi, } \] 

\noindent meaning\\

 \(~~ \frac{ 16 \sigma T^{3}}{3c}\overline{T} = \frac{ \omega _{B}^{2}}{k^{2}}\overline{ \rho }_{D}. \) \ \ \ \ \ \  \ \ \ \ \ \ \ \ \ \ \ \ \ \ \ \ \ \ \ \ \ \ \ \ \ \ \ \ \ \ \ \ \ \ \ \ \ \ \ \ \ \ \ \ \ \ \  \ \ \ \ \  \ \ \ \ (23)\\

\vspace{\baselineskip}
\noindent Now, dropping the relativistic  \( \frac{ \partial  \phi }{ \partial t} \)  term, Eq. (1) is\\

 \( \frac{d \rho _{B}}{dt}+  \rho _{B}\triangledown .~v_{B}~~= 0   \) \ \ \ \ \ \ \ \ \ \ \ \ \ \ \ \ \ \ \ \ \ \ \ \ \ \ \ \  \ \ \ \ \ \  \ \ \ \ \ \ \ \  \ \ \ \ \ \ \ \ \ \  \ \ \ \ \ \ \ (24)\\

\noindent and Eq. (5) is\\

 \[ ~~~~\frac{~dU_{R}}{dt}+ \left( U_{R}+p \right) \triangledown .~v_{B}=~0.~~~~~~~~~~~~~~~~~~~~~~~~~~~~~~~~~~~~~~~~~~~~~~~~~~~~~~~~~~~~~~~~~~ \left( 25 \right) ~~~~~~~~~~~    \] \\

\noindent Substituting for  \( \triangledown .~v_{B} \)  from Eqs. (24) to (25) yields

 \[ ~~~~\frac{1}{T}\frac{dT}{dt}~ = \frac{4}{3 \rho _{B}}\frac{d \rho _{B}}{dt},  \] 

\noindent and therefore\\

 \[~~~~ \frac{\overline{T}}{T}= \frac{4}{3}\frac{\overline{ \rho }_{B}}{ \rho _{B}}. \] 

\noindent And from Eq. (23),

 \[ ~~~~\frac{48 \sigma T^{4}}{9c}\frac{\overline{ \rho }_{B}}{ \rho _{B}} = \frac{ \omega _{B}^{2}}{k^{2}}\overline{ \rho }_{D}. \] 

\noindent Therefore, 

 \[~~~~ \overline{ \rho }_{B} =\frac{1}{4 \alpha }~\frac{ \omega _{B}^{2}}{k^{2}c^{2}}\overline{ \rho }_{D}, \] 

\noindent that is,\ \ \ \ \ \ \ \ \ \ \ \ \  \\
 \[~~~~ \overline{ \rho }_{B} \ll \overline{ \rho }_{D}. \] 

\noindent The baryons would be virtually unaffected by the collapse of the dark matter.\section{Conclusions}

  The eigenmode analysis here demonstrates the distinction between the baryon acoustic oscillation mode and the dark matter perturbation\ mode in the epoch between horizon crossing and recombination.  Baryon acoustic waves are fast moving, with a high frequency oscillating potential to which the dark matter barely reacts; on the other hand, the dark matter perturbation mode is a static, relatively\ slow\ gravitational instability.   The\ two\ modes\ are uncorrelated,  and the distinction between them is ignored in modern analyses,  where a constraint that couples the two modes is universally applied. 

Naoz and Barkana (2005), for instance, couple the two wave modes together by applying a constraint that the baryon and dark matter\ perturbations have the same spatial structure (i.e. a Fourier component with the same wavenumber and phase).  They\ then\ developed and numerically solved the resulting set of time-dependent differential equations for the perturbations. This dictates that the result will be a standing wave.   This technique is used by many papers – another example, using the Einstein-Boltzmann equations for photons, is Pan et al (2016), where their Eq. 8 explicitly shows the standing wave. 

Figure 3 of Naoz and Barkana indicates that under this constraint the fractional perturbations of the baryons and dark matter tend to equality after recombination due to the baryons falling into the coincident dark matter potential wells. This justifies the initial conditions assumed by large-scale cosmological simulations.

However, the analysis here shows that the baryon element of a dark matter perturbation wave is very small, and not oscillatory.  Hence, the standing wave in these calculations actually involves the coincidence of three separate waves – a dark matter perturbation wave and two baryon acoustic waves of equal magnitude travelling in opposite directions to create an oscillatory standing wave in phase with the dark matter wave.  This situation would arise if the quantum fluctuations present in the inflationary epoch, which are surmised to be responsible for the perturbations considered here,\ consisted\ solely of gravitational potential perturbations.   The effect of these gravitational perturbations would be to induce equal velocity perturbations in both the dark matter and baryons, leading to exactly this scenario. However, while the density perturbation in baryons and dark matter would be initially coincident, before recombination the density of the baryons would oscillate but that of the dark matter would not, so coincidence would occur only periodically.  Authors\ often make a statement to the effect that the potential generated by a dark matter density perturbation drives the oscillations in the baryons (Li et al. 2008 and  Eq. 8.18 of Dodelson 2003)\textit{, }but in fact an oscillatory baryon acoustic standing wave in phase with a dark matter\ perturbation  would be a consequence of initial conditions set up by gravitational fluctuations during inflation.

If\ the quantum fluctuations during inflation actually consist of density perturbations, rather than pure gravitational perturbations, then there is no reason why perturbations in the dark matter and baryons would be related.  Indeed, the differences highlighted here in the respective wave modes would make that highly unlikely.

On the assumption that all galaxies, and indeed galactic clusters, have their dark matter haloes, it could be argued that this justifies the assumption that dark matter and baryons should have coincident\ initial\ conditions\ for simulations.   However, this  would have difficulty accounting for the steeply falling rotation curve galaxies observed at z > $ \sim $ 1\  (Lang et al.\  2017 and Genzel et al. 2017).\  These early galaxies seem to lack dark matter\ haloes.  There are, however, very few, if any, steeply falling\ rotation curve galaxies in the local Universe – though recent stellar observations of a satellite galaxy of NGC1052 indicate that it may contain no dark matter (van Dokkum 2018).  If steeply falling rotation curves turn out to be present in all early galaxies,  and observations by the James Webb telescope may soon shed light on this, there would remain the requirement to explain how flat rotation curves develop as  galaxies age\ (see for instance Nelson 1988). \

Initial conditions for simulations should at least consider the possibility of dark matter and baryons not being coincident at z$ \sim $ 100.\  The\ possibility of propagating baryon acoustic oscillation (BAO) waves such as those that are deemed to create the $ \sim $ 150 Mpc  signature at z= 0 should also be considered. Tseliakhovich and Hirata (2010) propose that a relative velocity between dark matter and baryons due to the decrease in the sound velocity at recombination, going from radiation pressure to gas pressure,  will have a significant effect on the subsequent development of structure.   This effect is, however, second order in the perturbation amplitudes; the velocity and structural differences that would arise from the different evolution of dark matter perturbations and baryonic waves would be first order in the amplitudes.

Lastly, although the peaks in the power spectrum of the CMB anisotropies can be interpreted as standing waves in the finite length scale given by the sound horizon at the last scattering surface, for wavelengths much smaller than the sound horizon the evidence that there are standing\ waves is not so clear.  Indeed, the idea that BAO waves propagate away from any initial inflation-induced perturbation is backed up by the BAO signature at z=0 in the distribution of galaxies (Eisenstein et al, 2005).  Purely standing waves at smaller scales are not consistent with this.\  This, and the analysis here,  calls into question the assumption of standing baryonic waves in phase with dark matter in the analysis of the CMB anisotropies, as seen in Fig. 4 of Hu $\&$  Dodelson (2002), where fluctuations in the baryons are assumed to be driven by a potential created by perturbations in the dark matter.  It\ could be argued that the success of this model in reproducing the peaks in the power spectrum justifies this assumption, but Lopez-Corredoira  (2017) points out that matching the peaks can be achieved in a variety of ways not related to  \( \Lambda \)CDM.\\

\begin{acknowledgements}
     I thank Simon White for comments on the first draft of this paper.
\end{acknowledgements}

%

\begin{thebibliography}{}

\bibitem[]{} Boyd, T.J.M, Sanderson, J.J. ,1969, Plasma Dynamics, Thomas Nelson $\&$  Sons, London, p139.\\

\bibitem[]{} Dodelson, S., 2003, Modern Cosmology, Academic Press, San Diego, p84.\\

\bibitem[]{} Eisenstein, D.J.,  2005, Ap J, 633, 560.\\

\bibitem[]{} Genzel, R., 2017, Nature, 543, 397.\\

\bibitem[]{} Hu,\ W., Dodelson, S., 2002, Ann.Rev.Astr.Ap,  40, 171.\\

\bibitem[]{} Kaviraj, S. et al , 2017, MNRAS, 467, 4739.\\

\bibitem[]{} Lang, P. et al ,2017, ApJ, 840, 92.\\

\bibitem[]{} Li,\ B., 2008, Phys.Rev.D,  78, 4021.\\

\bibitem[]{} Lopez-Corredoira, M., 2017, Foundations of Physics, 47, 711.\\

\bibitem[]{} Naoz , S., Barkana, R., 2005, MNRAS, 362,1047.\\

\bibitem[]{} Nelson, A.H. , 1988, MNRAS, 233, 115.\  \\

\bibitem[]{} Pan, Z. et al, 2016, MNRAS, 459, 2513.\\

\bibitem[]{} Ruis-Granados, B., Battaner, E., Florido, E. ,2016, IAU Symposium 308, The Zeldovich Universe: Genesis and Growth of the Cosmic Web , p. 626.\\

\bibitem[]{} Schaye, J. et al, 2015, MNRAS, 446, 521. \\

\bibitem[]{} Silk, J., 1967, Nature, 215, 1155. \\

\bibitem[]{} Tseliakhovich, D., Hirata, C.M. , 2010, Phys.Rev.D, 82, 3520. \\

\bibitem[]{} van Dokkum, P., 2018, Nature, 555, 629.\\

\bibitem[]{} Vogelsberger, M. et al ,2014, Nature, 509, 177.\\

\end{thebibliography}
%

\end{document}